\newfont{\kreuz}{msbm10 scaled\magstep1}
\newfont{\Deutsch}{eufb10 scaled\magstep1}
\newfont{\deutsch}{eufb10}
\newfont{\schreib}{eusm10 scaled\magstep1}

\documentstyle[11pt]{article}

\renewcommand{\thefootnote}{\alph{footnote}}

\newcommand{\beqn}{\begin{equation}}
\newcommand{\eeqn}{\end{equation}}
\newcommand{\Om}{\Omega}
\newcommand{\Omp}{\Omega_{+}}
\newcommand{\Omm}{\Omega_{-}}
\newcommand{\Si}{\Sigma}
\newcommand{\Sip}{\Sigma_{+}}
\newcommand{\Sim}{\Sigma_{-}}
\newcommand{\Rpm}{R_{\pm}}
\newcommand{\Rp}{R_{+}}
\newcommand{\Rm}{R_{-}}
\newcommand{\Uqg}{\mbox{$U_{q}{\/(\mbox{\Deutsch g})}$}}

\begin{document}

\begin{titlepage}

\today          \hfill
\begin{center}
\hfill    LBNL-41209	 \\

\setcounter{footnote}{1}
\vskip .50in
\renewcommand{\thefootnote}{\fnsymbol{footnote}}

{\large \bf Quantum Algebra of the Particle Moving on the $q$\/-Deformed 
Mass-Hyperboloid}\footnote{This work 
was 
supported 
in
part by the Director, Office of Energy Research, Office of High Energy and 
Nuclear Physics, Division of High Energy Physics of the U.S. Department of 
Energy under Contract DE-AC03-76SF00098 and in part by the National Science
Foundation under grant PHY-95-14797}
\vskip .50in

Bogdan Morariu
\footnote{email address: bmorariu@lbl.gov}
\vskip .50in


{\em 	Department of Physics			\\
	University of California				\\
				and					
				\\
	Theoretical Physics Group			\\
	Lawrence Berkeley National Laboratory	\\
	University of California				\\
	Berkeley, California 94720}
\end{center}
\vskip .50in

\begin{abstract}
I introduce a reality structure on the Heisenberg double
of ${\rm Fun}_{q}(SL(N,{\rm C}))$ for $q$ phase, which for $N=2$ can 
be interpreted
as the quantum phase space of the particle on the $q$\/-deformed
mass-hyperboloid. This construction is closely related to the
$q$\/-deformation of the symmetric top. 
Finally, I conjecture that the above  real form 
describes
zero modes of certain non-compact WZNZ-models.
\end{abstract}
\end{titlepage}
\renewcommand{\thepage}{\roman{page}}
\setcounter{page}{2}
\mbox{ }

\vskip 1in

\begin{center}
{\bf Disclaimer}
\end{center}

\vskip .2in

\begin{scriptsize}
\begin{quotation}
This document was prepared as an account of work sponsored by the United
States Government. While this document is believed to contain correct
 information, neither the United States Government nor any agency
thereof, nor The Regents of the University of California, nor any of their
employees, makes any warranty, express or implied, or assumes any legal
liability or responsibility for the accuracy, completeness, or usefulness
of any information, apparatus, product, or process disclosed, or represents
that its use would not infringe privately owned rights.  Reference herein
to any specific commercial products process, or service by its trade name,
trademark, manufacturer, or otherwise, does not necessarily constitute or
imply its endorsement, recommendation, or favoring by the United States
Government or any agency thereof, or The Regents of the University of
California.  The views and opinions of authors expressed herein do not
necessarily state or reflect those of the United States Government or any
agency thereof, or The Regents of the University of California.
\end{quotation}
\end{scriptsize}

\vskip 2in

\begin{center}
\begin{small}
{\it Lawrence Berkeley National Laboratory is an equal opportunity employer.}
\end{small}
\end{center}

\newpage
\renewcommand{\thepage}{\arabic{page}}

\setcounter{page}{1}
\setcounter{footnote}{0}


\section{Introduction}

Monodromy matrices
representing the braid group~\cite{Kanie}, appearing in the WZNZ-model, 
suggested that hidden quantum groups exist  in these  theories.
Various approaches were used in an attempt to elucidate the
origin of these hidden quantum groups. 
In~\cite{CFT1,CFT2,CFT3,Faddeev1} using a 
Minkowski space-time lattice regularization,
it was shown by explicit
construction that the monodromies of the chiral components of the
WZNW-model with Lie group $G$ and the local field satisfy the commutation 
relations of the $q$-deformed cotangent bundle ${\rm T}^{*} G_{q}$.

However an apparent contradiction existed~\cite{CFT2,Faddeev1},
 since the deformation 
parameter in the WZNW-model must be  root of unity  $q=\exp(i\pi / k+h)$, 
where $k$ is the level of the affine-Lie algebra and $h$ is the 
dual Coxeter number, and this is incompatible with the compact 
form of the quantum group.

A solution to this problem was proposed in~\cite{Faddeev2}. The main 
idea is to drop the strong requirement that the reality structure be 
compatible with quantum group comultipication and only impose this 
requirement in the classical limit. Then a reality structure can be 
introduced, but not on the quantum group itself, but rather on the
quantum cotangent bundle.
               
However once the requirement of the compatibility of the 
reality structure with the comultiplication is dropped, one can introduce
more than one reality structure. In this paper I will introduce
one such reality structure inspired by a particular type of non-compact 
WZNW-model. 
See for example~\cite{Krysztof} for a 
list of various circumstances
under which this non-compact form occurs and also~\cite{Destri} where
the non-compact form of appears as the Euclidean section of the model.
These WZNW-models have the important property that the local field has 
the chiral decomposition $g=h h^{\dag}$ where $h$ is the chiral field
valued in $G$. Thus $g$ is a Hermitian positive defined matrix of unit
determinant. I will show that
\[
g^{\dag} = g
\]
is compatible with the algebra ${\rm T}^{*} G_{q}$ and extend the above
anti-involution to the whole algebra.
I emphasize that the reality structure introduced here
is similar to the one discussed in~\cite{Faddeev2} and is not related to 
the standard non-compact reality structure appearing in quantum groups
for $q$ phase, and which is compatible with comultiplication. 

For simplicity here I will not apply the reality structure 
directly in the WZNW-model, leaving this for a forthcoming paper,
 and instead I will just use it for the toy
model of~\cite{Faddeev1,Faddeev2}, which essentially contains all the
relevant degrees of freedom. These degrees of freedom
are described by the same algebra as 
in the compact case but with a different reality structure.

In Section~\ref{sec-algebra}, I give a short review of the quantum algebra
${\rm T}^{*} G_{q}$. I discuss the commutation relations for 
operators generating both left and right translations, since both forms
are necessary to define or to check the involutions presented in the next
sections. Section~\ref{sec-compact} briefly covers the reality
structure of~\cite{Faddeev2}. In section~\ref{sec-non-compact},  I 
present the main result of the paper, a reality structure corresponding
to a generalized  mass-hyperboloid configuration space and 
its associated $q$-deformed phase space. In 
Section~\ref{sec-mechanics}, I consider the simple quantum mechanical
system of~\cite{Faddeev2} and show its compatibility with the $*$-structure
introduced in the previous section. In the last Section I present some 
evidence for the relevance of this reality structure to the non-compact
WZNW-model.

\section{Review of the Algebra on ${\rm T}^{*} G_{q}$}
\label{sec-algebra}

In this section I present a brief review of 
the defining relations of the $q$\/-deformed
cotangent bundle~\cite{Faddeev1} also known as the Heisenberg double 
or as the 
smash product~\cite{Zumino,Zumino1}. The main purpose of this section is to
fix the notation. I will follow closely the presentation in~\cite{Faddeev2}
where a more detailed exposition can be found.

Let $G$ be the Lie group $SL(N,{\rm C})$, and sometimes 
for simplicity I will take  $G=SL(N,{\rm 2})$.
Most of the content of the paper can be easily  extended to 
arbitrary classical groups.
Now consider the quantum $\Rp$ matrix associated to the Lie group
$G$. This is a matrix depending on a parameter $q$\/ and
acting in the tensor product of two fundamental 
representations. For example the $\Rp$ of $SL(2,{\rm C})$ is the following
$4\times 4$ matrix
\[ \Rp = q^{-1/2} \left( \begin{array}{cccc}
		q & 0 & 0 & 0 \\
		0 & 1 & \lambda & 0 \\
		0 & 0 & 1 & 0 \\
		0 & 0 & 0 & q
				\end{array}
				\right)
\]
where $\lambda=q-q^{-1}$.
It is convenient to also use the $\Rm$\/ matrix defined as
\beqn
\Rm = P \Rp^{-1} P	\label{RmRp}
\eeqn
where $P$ is the permutation operator in the tensor space of the 
two fundamental representations
\[
P(a \otimes b) = b \otimes a	.
\]
Next I will define the quantum algebra ${\rm T}^{*} G_q$\/, the quantum
deformation of the cotangent bundle. 
Let $g$\/ and $\Om_{\pm}$\/ be matrices acting in the fundamental
representation of $G$\/. The $\Om_{\pm}$\/ matrices are upper and 
lower triangular matrices. In addition
the diagonal elements of $\Omp$\/ equal those of $\Omm^{-1}$\/.
${\rm T}^{*} G_q$\/ is the algebra generated by $g$\/ and $\Om_{\pm}$\/ and 
satisfying 
 the following set of relations divided for
convenience into  three groups
\beqn
\Rpm g^{1} g^{2} = g^{2} g^{1} \Rpm   \label{Rgg}
\eeqn
\begin{eqnarray}	
\Rpm  \Omp^{1} \Omp^{2} = \Omp^{2} \Omp^{1} \Rpm    \nonumber	\\
\Rpm  \Omm^{1} \Omm^{2} = \Omm^{2} \Omm^{1} \Rpm \label{RLLplus} \\
\Rp \Omm^{1} \Omm^{2} = \Omm^{2} \Omp^{1} \Rp  \nonumber \\
\Rm \Omm^{1} \Omp^{2} = \Omp^{2} \Omm^{1} \Rm  \nonumber
\end{eqnarray}
\begin{eqnarray}	
\Rp  \Omp^{1} g^{2} = g^{2} \Omp^{1}  \label{RLgplus} \\
\Rm  \Omm^{1} g^{2} = g^{2} \Omm^{1}.  \nonumber
\end{eqnarray}
All the above relations are operator matrices acting in the tensor product
of two fundamentals, and the superscript indicates on which factor 
the respective matrix acts. The $R$ matrices without any superscript
act in both spaces.
One can show that the quantum determinant of the 
matrices $g$ and $\Om_{\pm}$ is central and can be set equal to one
\[
{\textstyle \det_{q}} (g) = {\textstyle\det_{q}} (\Om_{\pm}) = 1.
\]
For the $SL(N,{\rm C})$ groups these are all the relations, 
while for the other
classical groups additional relations, for example orthogonality relations,
have to be imposed.
Noto also that, unlike~(\ref{Rgg})(\ref{RLLplus}),  the 
relation~(\ref{RLgplus}) is not homogeneous in $\Rpm$ thus the normalization
of $\Rpm$ is important.

The above relations are not independent. For example the $\Rm$\/ relations
can be obtained from the $\Rp$\/ relations using~(\ref{RmRp}) and 
\beqn
X^{2}=P X^{1} P.	\label{PXP}
\eeqn

The subalgebra generated by the matrix elements of $g$\/ with 
relations~(\ref{Rgg}) is in fact a Hopf algebra 
denoted ${\rm Fun}_{q}(G)$\/ and
represents a deformation of the Hopf algebra of function on the $G$ Lie
group~\cite{FRT}. 
Also, the subalgebra
generated by $\Om_{\pm}$\/ with relations~(\ref{RLLplus})  is 
a quasitriangular Hopf algebra called
the quantum universal enveloping algebra~\cite{Drinfeld,Jimbo,FRT}, and is 
denoted $\Uqg$\/ 
where the \mbox{\Deutsch g} in the brackets is the Lie algebra of
the Lie group $G$\/. For example 
the coproduct of ${\rm Fun}_{q}(G)$\/ on the matrix 
elements of $g$\/ is given by
\beqn
\triangle (g) = g \dot{\otimes} g ,
\eeqn
where the dot means multiplication in matrix space. Similarly the coproduct
in $\Uqg$\/ on the matrix elements $\Om_{\pm}$\/ reads
\beqn
\triangle (\Om_{\pm}) = \Om_{\pm} \dot{\otimes} \Om_{\pm} .
\eeqn
On the other hand ${\rm T}^{*} G_{q}$ is not a Hopf algebra. We emphasize 
this, since
there is a related algebra, the Drinfeld double, which has the same 
generators but different mixed relations and is a Hopf algebra.

The mixed relations~(\ref{RLgplus}) describe how to combine
 the above subalgebras into the larger algebra ${\rm T}^{*} G_{q}$.
They appear as commutation relations in~\cite{Faddeev1,Zumino,Zumino1} 
but in an abstract form as the pairing
of dual Hopf algebras  they were already
present in~\cite{FRT}.

One can relate the $\Om_{\pm}$ with the more traditional Drinfeld-Jimbo
generators. For example for the $SL(2,{\rm C})$  group we can  write the
matrix elements of $\Om_{\pm}$ as~\cite{FRT}
\beqn	\Omp = \left(	\begin{array}{cc}
	q^{-H/2} & q^{-1/2} \lambda X_{+}	\\
	0	& q^{H/2}
			\end{array}
					\right) ,~
	\Omm = \left(	\begin{array}{cc}
	q^{H/2}		& 	0	\\
	-q^{1/2} \lambda X_{-}	& q^{-H/2}
			\end{array}
					\right)	.
\eeqn
Usinq the $\Rp$ matrix above it can be shown by direct computations 
that the generators
$H,X_{\pm}$\/ satisfy the Jimbo-Drinfeld relations~\cite{Drinfeld,Jimbo}
\beqn
[H,X_{\pm} ]= \pm 2 X_{\pm}, ~~ [X_+,X_-]=\frac{q^{H}-q^{-H}}{q-q^{-1}}
\eeqn
defining the universal enveloping algebra ${\cal U}_q(sl(2,{\rm C}))$. 
Similar
relations also exist for higher rank groups~\cite{FRT} and can be 
thought of as connecting the Cartan-Weyl and Chevalley bases.

It is also convenient to combine $\Om_{\pm}$ into a single 
matrix~\cite{RS}
\beqn
\Om = \Omp \Omm^{-1}.
\eeqn
In terms of these generators all the relations~(\ref{RLLplus}) 
and~(\ref{RLgplus}) collapse to
\begin{eqnarray}	
\Om^{1} \Rm^{-1} \Om^{2} \Rm &=& \Rp^{-1} \Om^{2} \Rp \Om^{1} \label{RLL} \\
\Rm  g^{1} \Om^{2} &=& \Om^{2} \Rp g^{1}. \nonumber
\end{eqnarray}
These forms of the commutation relations are especially useful when we deal 
with the commutation relations only, but the coproduct of $\Om$ cannot 
in general be given in an explicit form.

The commutation relations~(\ref{Rgg})(\ref{RLL}) are exactly those
satified by the local field and the monodromy of the left (or right)
chiral component of the affine current~\cite{CFT1,CFT2,CFT3}.

Following~\cite{Faddeev2} we also introduce an equivalent description
of the quantum algebra using operators generating right translations. First 
let 
\[
\Si = g^{-1} \Om g ,
\]
and then introduce a triangular decomposition of $\Si$ into $\Si_{\pm}$\/ 
\beqn
\Si = \Sip \Sim^{-1}
\eeqn
similar to the decomposition of $\Om$ into $\Om_{\pm}$.
One can check that the matrix elements of $\Om$\/ and $\Si$\/ commute.
To make the picture more symmetric also introduce a new matrix $h$\/ by
\beqn
h = \Si_{\pm}^{-1} g^{-1} \Om_{\pm}.	\label{h-of-g}
\eeqn
Now we can use either pair $(g,\Om)$\/ or $(h,\Si)$\/ to describe the 
algebra${\rm T}^{*} G_{q}$.

The defining relations satisfied by $h$\/ and $\Si$\/ are~\cite{Faddeev2}
\begin{eqnarray}	
\Rpm h^{1} h^{2} &=& h^{2} h^{1} \Rpm  		   \nonumber	   \\
\Sip^{1} \Sip^{2} \Rpm &=& \Rpm  \Sip^{2} \Sip^{1}   \label{RSSplus} \\
\Sim^{1} \Sim^{2} \Rpm &=& \Rpm  \Sim^{2} \Sim^{1}   \nonumber       \\
\Sim^{1} \Sip^{2} \Rp  &=& \Rp   \Sim^{2} \Sim^{1}   \nonumber       \\
\Sip^{1} \Sim^{2} \Rm  &=& \Rm   \Sim^{2} \Sip^{1}   \nonumber	   \\
h^{1} \Sip^{2} &=& \Sip^{2} \Rm h^{1}		   \nonumber       \\
h^{1} \Sim^{2} &=& \Sim^{2} \Rp h^{1}.		   \nonumber
\end{eqnarray}
One can check directly the consistency of~(\ref{RSSplus}) with the original
relations.

\section{Real Form for the $q$\/-Deformed Symmetric Top}
\label{sec-compact}

For a large number of applications the variable $q$ is a phase. In this 
case the $\Rpm$ matrices satisfy
\beqn
\Rp^{\dag} = \Rm.  \label{R-dagger}
\eeqn
If we require a reality structure for $g$ compatible with the Hopf 
algebra structure i.e.  
\[
\triangle \circ * = ( * \otimes * ) \circ \triangle
\]
and use~(\ref{R-dagger}) we obtain a
non-compact quantum group. For example if $G= SL(N,{\rm C})$ we obtain
${\rm Fun}_{q}(SL(N,{\rm R}))$.  

However sometimes in the same application 
we  are interested in the compact form of the group. This apparent 
contradiction can be  resolved~\cite{Faddeev2} by dropping the above 
requirement for a  Hopf $*$\/-structure. Instead one defines an 
anti-involution on the larger algebra ${\rm T}^{*} G_{q}$
\begin{eqnarray}
 \Om_{\pm}^{\dag} &=& \Om_{\mp}	\label{Omdag}  \\
	g^{\dag} &=& h.  \label{g-dagger}
\end{eqnarray}

It is straightforward~\cite{Faddeev2} 
to check the compatibility of this anti-involution
with the quantum 
algebra~(\ref{Rgg})(\ref{RLLplus})(\ref{RLgplus})(\ref{RSSplus}).
Note that~(\ref{Omdag}) does not define a Hopf $*$\/-structure on 
$\Uqg$, and~(\ref{g-dagger}) does not close on
${\rm Fun}_{q}(G)$ since the definition  of $h$ includes generators
of $\Uqg$. In the classical limit~(\ref{g-dagger}) 
reduces to $g^{\dag}=g^{-1}$\/ and~(\ref{Omdag}) becomes compatible 
with the coproduct. This is due to the fact that the coproduct is
cocommutative at$q=1$.

\section{Real Form for the $q$\/-Deformed Hyperboloid}
\label{sec-non-compact}

This section contains the main result of the paper, an
anti-involution on the deformed cotangent bundle when $q$ is a phase.
Like the anti-involution of the previous section, it does not originate from 
a Hopf $*$-structure on one of the Hopf subalgebras. 
The defining relations of the anti-involution are
\begin{eqnarray}
	g^{\dag} &=& g		\label{g-dag}		\\
	\Om_{\pm}^{\dag} &=& \Si_{\mp}^{-1}.   \label{Om-dag}
\end{eqnarray}
Alternatively the second relation can be written as
\beqn
	\Om^{\dag} = \Si = g^{-1} \Om g.	\label{Om-dag2}
\eeqn
It is quite obvious that~(\ref{g-dag}) is not compatible with the coproduct,
i.e. $g$ should not be considered a ``group element''.
I will not give a complete proof of the consistency of the anti-involution 
with the algebra relations~(\ref{Rgg})(\ref{RLLplus})(\ref{RLgplus}).
Instead I will just give a sample computation leaving the rest for the 
interested reader.

Applying the involution on the $\Rp$ relation~(\ref{Rgg}) 
and using~(\ref{R-dagger}) we have
\[
(g^{2})^{\dag}(g^{1})^{\dag} \Rm = 
\Rm (g^{1})^{\dag}(g^{2})^{\dag}.
\]
Moving the $\Rm$ matrices to the other side and using~(\ref{RmRp}) we 
obtain
\[
\Rp (g^{1})^{\dag} (g^{2})^{\dag} =
(g^{2})^{\dag} (g^{1})^{\dag} \Rp,
\]
thus it is consistent with the algebra relations~(\ref{Rgg}) to impose
$g^{\dag} =g$.

As another example, take the hermitian conjugate of 
the following relation
\beqn
\Rp  \Omp^{1} \Omp^{2} = \Omp^{2} \Omp^{1} \Rp .
\eeqn 
Using~(\ref{Om-dag}) we obtain
\[
(\Sim^{2})^{-1} (\Sim^{1})^{-1} \Rm = \Rm (\Sim^{1})^{-1} (\Sim^{2})^{-1}
\]
which can be rewritten after multiplication by some inverse matrices as
\[
\Rm \Sim^{2} \Sim^{1} = \Sim^{1} \Sim^{2} \Rm.
\]
This is just one of the equations in~(\ref{RSSplus}). 

Similarly applying the above involution on the first relation 
in~(\ref{RLgplus}) we obtain
\[
g^{2} (\Sim^{1})^{-1} \Rm = (\Sim^{1})^{-1} g^{2}
\]
\[
\Sim^{1} g^{2} = g^{2} \Rm^{-1} \Sim^{1}.
\]
This is equivalent  using~(\ref{RmRp}) and~(\ref{PXP}) to
\[
\Sim^{2} g^{1} = g^{1} \Rp \Sim^{2},
\]
and after eliminating $g$\/ using~(\ref{h-of-g}) we get
\[
\Sim^{2} \Omm^{1} (h^{1})^{-1} (\Sim^{1})^{-1} = 
\Omm^{1} (h^{1})^{-1} (\Sim^{1})^{-1} \Rp \Sim^{2}.
\]
Furthermore using~(\ref{RSSplus}) to commute the $\Si$ matrices we have
\[
 \Sim^{2} \Omm^{1} (h^{1})^{-1} =
\Omm^{1} (h^{1})^{-1} \Sim^{2} \Rp 
\]
and since $\Om$\/ and $\Si$\/ commute with each other we  finally obtain
\[
h^{1} \Sim^{2} = \Sim^{2} \Rp h^{1}	
\]
which is again one of the relations in~(\ref{RSSplus}). All the other
relations can be checked in a similar fashion.

Finally I will explain the terminology used in the title of this section.
Consider first for simplicity the $SL(2,{\rm C})$ case. 
In the undeformed case a 
$2 \times 2$ hermitian matrix of unit determinant defines the unit mass
hyperboloid in Minkowski space. For simplicity I will only consider 
one connected component of the manifold, for example the future mass
hyperboloid. For a general group $G$ this can be achieved 
by restricting to positive definite matrices. In the deformed case 
we consider Hermitians $g$ matrices of unit quantum determinant.

\section{Quantum Mechanics on the $q$-Deformed Hyperboloid}
\label{sec-mechanics}

In~\cite{Faddeev1} Alekseev and Faddeev showed  that 
the ${\rm T}^{*} G_{q}$ quantum algebra
is a $q$\/-deformation of the algebra of functions 
on the cotangent bundle of 
the Lie group $G$. In~\cite{Faddeev2} they considered the following 
simple Lagrangian written in first order formalism
\beqn
{\cal L} = {\rm Tr} (\omega \dot{g} g^{-1} - \frac{1}{2}\omega^{2}) .
\label{Lag}
\eeqn
Here $G$ is considered without specifying its real form. The Lagrangian 
has a chiral symmetry $G \times G$ 
\[
g \rightarrow u g v^{-1},~~\omega \rightarrow u \omega v^{-1},~~u,v \in G.
\label{chiral}
\]
The second order form of the 
Lagrangian has the form of a non-linear sigma model in $(0,1)$ dimensions
\beqn
{\cal L} = \frac{1}{2} {\rm Tr} ( \dot{g} g^{-1}~\dot{g} g^{-1}). 
\label{Lag2ord}
\eeqn
The equations of motion 
\[
\dot{g}=\omega g ,~~ \dot{\omega} = 0
\]
can be integrated to give the time evolution
\begin{eqnarray}	
\omega(t) &=& \omega(0)  \nonumber \\
g(t) &=& \exp(\omega t) ~g(0). \nonumber
\end{eqnarray}	

The real form corresponding to the compact group discussed 
in~\cite{Faddeev2} is
\beqn
g^{\dag} = g^{-1},~~\omega^{\dag} = - \omega. \label{real0}
\eeqn
For $G=SL(2,{\rm C})$, $g$\/ becomes unitary and 
the Lagrangian~(\ref{Lag}) describes the
classical dynamics of the symmetric top. Equivalently, it describes
the motion on a constant curvature $S^{3}$. This can be seen using the
chiral symmetry~(\ref{chiral}) of the Lagrangian, which under the 
conditions~(\ref{real0})  is restricted to the 
$SU(2) \times SU(2)  \sim SO(4)$ subgroup, or by direct computation of 
the metric in the kinetic term of~(\ref{Lag2ord}).

Instead, we consider the following reality structure
\beqn
g^{\dag} = g,~~ \omega^{\dag} = g^{-1} \omega g  \label{real1}
\eeqn
which, following from  the discussion at the end of the previous, section 
defines the phase space of a particle moving on the mass-hyperboloid.
The reality structure~(\ref{real1}) requires  $u^{\dag} = v^{-1}$
thus restricting the chiral symmetry of the 
Lagrangian to one independent $SL(2,{\rm C})$ subgroup which is simply
the Lorentz group that leaves
the mass hyperboloid invariant. The metric on the hyperboloid is just the 
induced metric from Minkowski space, and again this can be obtained by 
direct computation or using the above invariance under the Lorentz group.

One can check that the equations of motion preserve 
both reality structures~(\ref{real0}) and~(\ref{real1}).
What we learn from this simple 
example is that one can find rather different physical
systems that 
will have the same Poisson brackets and thus quantum algebras if their 
respective Lagrangians have the same form, differing only through their
reality structures.

In~\cite{Faddeev2} a $q$\/-deformation of the above system was introduced.
The model has a discrete time dynamics, with the time labelled by an integer
$n$. The following evolution equations
\begin{eqnarray}
\Om(n) &=& \Om(0)   \label{eom} \\
g(n) &=& \Om^{n} g(0) \nonumber
\end{eqnarray}
were shown in~\cite{Faddeev2} to preserve the quantum 
algebra~(\ref{Rgg})(\ref{RLLplus})(\ref{RLgplus}) and in addition, the 
reality structure discussed in Section~\ref{sec-compact}. 

I will now show 
that they also preserve the reality structure introduced in 
Section~\ref{sec-non-compact}.
Assuming that for $n=0$ the reality 
structure is given by~(\ref{g-dag}) and~(\ref{Om-dag2}) 
\[
g^{\dag}(0) = g(0),~~\Om^{\dag}(0)=g^{-1}(0) ~\Om(0)~ g(0)
\]
for arbitrary $n$ we have
\[
g^{\dag}(n) = g^{\dag}(0) (\Om^{\dag}(0))^{n}=
g(0) (g^{-1}(0)\Om ()) g(0))^{n}=
\Om^{n}(0) g(0) = g(n).
\]
Similarly we have for $\Om (n)$
\[
\Om^{\dag}(n) = \Om^{\dag}(0) = g^{-1}(0) \Om(0) g(0) = 
g^{-1}(n) \Om(0) g(n) =   g^{-1}(n) \Om(n) g(n).
\]
Thus  the equations of motion~(\ref{eom}) and the reality structure of the 
previous Section
define the $q$\/-deformation of the dynamics of a 
particle on the unit mass hyperboloid. 

\section{Concluding Remarks}
\label{sec-conclusion}

I conclude by briefly applying the reality structure  to the lattice
regularized WZNW-model and checking its compatibility with periodic 
boundary conditions. Using the notation in~\cite{CFT1} 
let the lattice have $N$ points, and denote the local 
fields by $g_{i},~i=1\ldots N$. For periodic boundary conditions we 
identify $i$ and $i+N$. Let $M_{L}$ and $M_{R}$ be  the monodromies
of the left and right affine currents. 
The algebra satisfied by $(g,M_{L},M_{R})$ is exactly the 
algebra of ${\rm T}^{*} G_{q}$ for the generators $(g,\Om,\Si)$.
Here I used the remark of the previous Section that the compact and
non-compact WZNW-models have the same algebra since their respective
Lagrangians coincide.
The monodromies  can be used to relate 
the fields $g_{0}$ and $g_{N}$
\[
g_{N} = M_{L} g_{0} M_{R}^{-1}
\]
If we require $g_{0}^{\dag} = g_{0},~~M_{L}^{\dag}=M_{R}$, which is 
just the reality structure of Section~\ref{sec-non-compact}, we have
\[
g_{N}^{\dag}= (M_{R}^{-1})^{\dag} g_{0} M_{L}^{\dag} =
M_{L}^{-1} g_{0} M_{R}=g_{-N} = g_{N}.
\]
In the last step I used the lattice periodicity. Thus we see that the
reality structure is compatible with periodic boundary conditions.
A more detailed investigation of the implications of this reality structure
for the WZNW-model will be presented in an upcoming paper.

\section*{Acknowledgements}

I would like to thank Professor Bruno Zumino for many helpful discussions
and suggestions. I would also like to thank Paolo Aschieri for valuable 
comments. This work was supported in part by 
the Director, Office of Energy Research, Office of High Energy and Nuclear
Physics, Division of High Energy Physics of the U.S. Department of Energy
under Contract DE-AC03-76SF00098 and in part by the National Science 
Foundation under grant PHY-95-14797.


\begin{thebibliography}{9}

\bibitem{Kanie} A. Tsuchiya, Y. Kanie {\em Vertex Operators in the 
Conformal Field Theory on ${\rm P}^{1}$\/ and Monodromy 
Representations of the Braid Group}, Letters in Math. Phys. 13 
(1097) 303-312


\bibitem{CFT1} A. Yu. Alekseev, L.D. Faddeev, 
M. A. Semenov-Tian-Shansky, A. Volkov
{\em The Unraveling of the Quantum Group Structure in the WZNW
theory}\/, Preprint CERN-TH-5981/91, January1991

\bibitem{CFT2} A. Yu. Alekseev, L.D. Faddeev, 
M. A. Semenov-Tian-Shansky 
{\em Hidden Quantum Groups Inside Kac-Moody Algebra}\/, Commun. 
Math. Phys. 149  (1992)  335-345

\bibitem{CFT3} L.D. Faddeev, {\em From Integrable Models to 
Conformal Field Theory via Quantum Groups}\/, Integrable Systems,
Quantum Groups, and Quantum Field Theory, L. A. Ibort, M. A. 
Rodr\'{i}quez (eds.)

\bibitem{Faddeev1} A. Yu. Alekseev, L.D. Faddeev
{\em $({\rm T}^{*} G)_{t}$\/: A Toy Model for Conformal Field Theory},\/ 
Commun. Math. Phys. 141  (1991)  413-422

\bibitem{Faddeev2} A. Yu. Alekseev, L.D. Faddeev
{\em An Involution and Dynamics for the $q$\/-Deformed Quantum 
Top}, Preprint hep-th/9406196, June 1994

\bibitem{Krysztof} K. Gaw\c{e}dzki, {\em Non-Compact WZW 
Conformal Field Theories}\/, Preprint hep-th/9110076, October 1991

\bibitem{Destri} C. Destri, H. J. De Vega {\em On The Connection 
Between The Principal Chiral Model and the Multiflavour Chiral 
Gross-Neveu Model}\/, Phys. Lett. B 201 (1988) 245-250

\bibitem{Zumino} B. Zumino, {\em Introduction to the Differential Geometry
of Quantum Groups}, K. Schm\"{u}dgen (Ed.), Math. Phys. X, Proc. X-th IAMP 
Conf. Leipzig (1990), Springer-Verlag (1991)

\bibitem{Zumino1} B. Zumino, {\em Differential Calculus on Quantum Spaces
and Quantum Groups}, XIX ICGTMP, M. O., M. S. and J. M. G. (Ed.),
CIEMAT/RSEF, Madrid, vol. 1 (1993)~4

\bibitem{FRT} L.D. Faddeev, N. Yu. Reshetikhin, L. A. Takhtajan 
{\em Quantization of Lie Groups and Lie Algebras}, Alg. i Anal. 1 
(1989) 178

\bibitem{Drinfeld} V. G. Drinfeld, {\em Quantum Groups}, ICM MSRI, 
Berkeley  (1986) 798-820
 
\bibitem{Jimbo} M. Jimbo {\em A $q$\/-Difference Analogue 
of ${\rm U}(g)$\/
of the Yang-Baxter Equation}\/,
Lett. Math. Phys. 10 (1985) 63-69

\bibitem{RS} N. Yu. Reshetikhin, M. A. Semenov-Tian-Shansky, {\em  
Quantum R-matrices and Factorization Problems}, JGP. Vol. 5, 
nr.~4 (1988) 534-550

\end{thebibliography}
\end{document}